\documentclass[12pt,epsf]{iopart}

\usepackage{pstricks}
\usepackage{pst-coil}
\usepackage{amssymb}
\usepackage{graphics,epsf}

\begin{document}

\title{Odd-parity perturbations of self-similar Vaidya space-time.}
\author{Brien C. Nolan}
\address{School of Mathematical Sciences, Dublin City
University, Glasnevin, Dublin 9, Ireland.}
\eads{\mailto{brien.nolan@dcu.ie}}
\begin{abstract}
We carry out an analytic study of odd-parity perturbations of the
self-similar Vaidya space-times that admit a naked singularity. It
is found that an initially finite perturbation remains finite at the
Cauchy horizon. This holds not only for the gauge invariant metric
and matter perturbation, but also for all the gauge invariant
perturbed Weyl curvature scalars, including the gravitational
radiation scalars. In each case, `finiteness' refers to Sobolev
norms of scalar quantities on naturally occurring spacelike
hypersurfaces, as well as pointwise values of these quantities.
\end{abstract}
\submitto{CQG}  \maketitle


\newtheorem{assume}{Assumption}
\newtheorem{theorem}{Theorem}
\newtheorem{prop}{Proposition}
\newtheorem{corr}{Corollary}
\newtheorem{lemma}{Lemma}
\newtheorem{definition}{Definition}
\newtheorem{remarkthe}{Remark}[theorem]
\newtheorem{remarkcorr}{Remark}[corr]
\newtheorem{remarklem}{Remark}[lemma]
\newcommand{\be}{\begin{equation}}
\newcommand{\ee}{\end{equation}}
\newcommand{\bq}{\begin{eqnarray}}
\newcommand{\eq}{\end{eqnarray}}
\newcommand{\by}{\begin{eqnarray*}}
\newcommand{\ey}{\end{eqnarray*}}
\newcommand{\so}{{\Sigma_i}}
\newcommand{\sco}{{\cal{O}}}
\newcommand{\ch}{{\cal{H}}}
\newcommand{\chp}{{\ch^+}}
\newcommand{\ce}{{\cal{E}}}
\newcommand{\cf}{{\cal{F}}}
\newcommand{\cm}{{\cal{M}}}
\newcommand{\cmt}{\tilde\cm}
\newcommand{\pnc}{{\cal{N}}}
\newcommand{\bphi}{{\tilde{\phi}}}
\newcommand{\hE}{{\hat{E}}}
\newcommand{\kp}{\kappa}
\newcommand{\lam}{\lambda}
\newcommand{\vv}{\vec{\varphi}}
\newcommand{\vj}{\vec{\jmath}}
\newcommand{\tmu}{\tilde{\mu}}
\newcommand{\bmt}[1]{\mbox{\boldmath $#1$}}
\def\fin{\hfill \rule{2.5mm}{2.5mm}\\ \vspace{0mm}}
\def\bs{\rule{2.5mm}{2.5mm}}

\section{Introduction and summary: naked singularities in self-similar collapse.}
In the standard picture of gravitational collapse, the implosion
that results from the instability of the collapsing object leads to
the formation of a black hole horizon prior to the formation of the
inevitable singularity \cite{mtw}. However, it is known that models
exist where the `horizon before singularity' order is not followed,
and the singularity that results is visible to external observers.
There are different reasons why the existence of such naked
singularities are an undesirable feature of spacetime: on a
fundamental level, they are accompanied by Cauchy horizons leading
to a breakdown in predictability of physical laws. On a physical
level, the possibility arises that naked singularities may be the
source of infinite (destructive) amounts of energy. The Cosmic
Censorship Hypothesis (CCH) of Penrose seeks to guard against naked
singularities. In rough terms, this hypothesis asserts that naked
singularities cannot form in realistic gravitational collapse (there
are rigorous mathematical formulations of the hypothesis; see for
example \cite{wald}). Thus those models which give rise to naked
singularities must be unrealistic in some way. Nevertheless, models
admitting naked singularities provide probes of the CCH, and studies
of naked singularities have influenced the development of exact
statements of the hypothesis. There is also some hope that studies
of spacetimes admitting naked singularities will shed light on how a
general proof of the hypothesis might arise. Of course one must also
keep in mind the possibility that such a model {\em cannot} be ruled
out, and that naked singularities must be considered to be genuine
astrophysical objects.

There are at least three ways in which spacetime models admitting
naked singularities can be ruled out as providing genuine
counterexamples to the CCH. First, the matter model used may be
considered to be inappropriate to the description of gravitational
collapse on the smallest scale. This is understood to be the case,
for example, in fluid models: the singularities that result are
ascribed to a breakdown of the matter model rather than a
gravitational pathology \cite{rendall}. Indeed careful statements of
the hypothesis insist that the matter model used must be such that
it does not develop singularities in flat spacetime \cite{wald}.

The second way is to demonstrate that the model that includes a
naked singularity is unstable in the following way. One shows that
a small perturbation of the initial data for the spacetime gives
rise to a spacetime that does not admit a naked singularity. This
approach has been used by Christodoulou to show that naked
singularities forming in the self-similar collapse of a
spherically symmetric massless scalar field are unstable
\cite{christo}.

Finally, one looks for instability of the Cauchy horizon
associated with the naked singularity. In this scenario, the model
admitting a naked singularity is a non-generic member of a class
of spacetimes which instead give rise to a null singularity
marking the end of the spacetime rather than a problematic
horizon. This situation holds a the inner (Cauchy) horizon of
charged or rotating black holes \cite{brady}.

In this paper, we deal with a 1-parameter family of spherically
symmetric self-similar spacetimes that admit a naked singularity:
the self-similar Vaidya spacetimes. We will study stability of the
associated Cauchy horizon.

The question of whether or not these spacetimes provide a serious
challenge to the CCH can be answered immediately in the negative.
The matter model is null dust, which always forms singularities in
flat spacetime. However, there are many classes of spherically
symmetric self-similar spacetimes admitting naked singularities
that cannot be ruled out on this basis. By studying stability of
the Cauchy horizon of Vaidya spacetime, we will provide a template
for the study of the same issue in more realistic spacetimes
(e.g.\ self-similar Lema\^{\i}tre-Tolman-Bondi, perfect fluids,
Einstein-Klein-Gordon, Einstein-$SU(2)$). There are also other
reasons for studying perturbations of Vaidya spacetime. This
spacetime is used to model the late stages of stellar collapse in
which radiative emission dominates. Studying stability of the
spacetime provides information with regard to the effectiveness of
this model.

A spacetime is said to be self-similar if it admits a homothetic
Killing vector field, i.e.\ a vector field $\vec{X}$ satisfying
\[ {\cal{L}}_{\vec{X}}g_{ab}=2g_{ab}.\] (The choice of non-zero
constant on the right hand side is arbitrary, and it should be
noted that some authors would refer to $\vec{X}$ as defined here
as a proper homothetic Killing vector field, or to the associated
symmetry as type-1 self-similarity.) See \cite{carr-coley} for an
overview of the important role of self-similarity in General
Relativity.

The line element of a spherically symmetric spacetime $(M,g)$ can
always be written in the form \be
ds^2=2F(v,r)e^{P(v,r)}dv^2+2e^{P(v,r)}dvdr+r^2d\Omega^2,\label{lel}\ee
where $d\Omega^2$ is the standard line element on the unit
2-sphere. The coordinate $v$ is an advanced Bondi coordinate.
Taking this null coordinate to increase into the future, it labels
past null cones of the axis $r=0$. The line element above
maintains its form under the relabelling $v\to V(v)$.
Self-similarity holds if and only if $F(v,r)=G(t)$ and
 $P(v,r)=\psi(t)$ for some functions $G,\psi$ and where $t=\frac{v}{r}$:
\be
ds^2=-2G(t)e^{\psi(t)}dv^2+2e^{\psi(t)}dvdr+r^2d\Omega^2.\label{sslel}\ee
The homothetic Killing vector field is
$\vec{X}=v\frac{\partial}{\partial v}+r\frac{\partial}{\partial r}$.

We will use the coordinates $t=v/r$ and a rescaled radial coordinate
$x$ defined by $r=e^x$. Then the line element reads \be
ds^2=e^{2x}\left\{-2Ge^\psi
dt^2+2e^\psi(1-2tG)dtdx+2te^\psi(1-tG)dx^2+d\Omega^2\right\}.\label{selfsimlel}\ee

A general description of spherically symmetric spacetimes modelling
gravitational collapse has been given in \cite{nolan-waters1} and
\cite{nolan-scalar}. This can be done without specifying the matter
model. Here, we restrict ourselves to the following crucial points,
proven in \cite{nolan-waters1}. The second result gives necessary
and sufficient conditions for the singularity that necessarily forms
at $(v,r)=(0,0)$ to be naked.

\begin{prop} The surface $t=t_c$ constant is spacelike
(respectively, null, timelike) if and only if $t_c(1-t_cG(t_c))>0$
(respectively, $=0,<0$).
\end{prop}

This gives rise to the moniker `similarity horizon' for null
hypersurfaces of the form $t=t_c$.

\begin{prop}
Suppose that the spacetime $(M,g)$ with line element (\ref{sslel})
\begin{itemize}
\item[(i)] satisfies the Einstein equation;
\item[(ii)] has energy-momentum tensor satisfying the null energy
condition;
\item[(iii)] is regular to the past of the scaling origin
$\sco=(v,r)=(0,0)$, where $v$ is scaled to measure proper time on
the regular axis $\{v<0,r=0\}$.
\end{itemize}
Then there exists a future-pointing outgoing radial null geodesic
with past endpoint on $\sco$ if and only if there is a positive
solution of the equation $1-tG(t)=0$. Furthermore, if $t_1$ is the
smallest such positive root, then the surface $t=t_1$ is the
Cauchy horizon of the spacetime.\fin
\end{prop}

We have not defined `regularity to the past of $\sco$'; it
suffices to note that this is a well-defined concept, including
limiting behaviour of the metric at the regular axis and the
absence of trapped surfaces in $v<0$. Note also that we can
characterise the Cauchy horizon as being the first similarity
horizon to the future of the scaling origin. The past null cone of
the scaling origin is a similarity horizon given by $t=0$.

In the following section, we describe the geometry of Vaidya
spacetime, the subject of our analysis, concentrating on the
self-similar case. In Section 3, we describe the gauge-invariant
approach to perturbations of spherically symmetric spacetimes given
by Gerlach and Sengupta \cite{GS}. We show that for odd-parity
perturbations (see below), the matter perturbation is completely and
explicitly determined by an initial data function $\mu_0$. The
remaining perturbation quantities are determined through a single
gauge-invariant scalar $\Pi$, which satisifies an inhomogeneous wave
equation with source term depending on $\mu_0$. We give existence
and uniqueness results, and show that, subject to the specification
of regular initial data on a slice $t=t_i\in(0,t_1)$, the function
$\Pi$ and its first partial derivatives remain finite up to and at
the Cauchy horizon $t=t_1$. Thus there is no instability at the
level of the metric or the matter perturbation. There remains the
possibility that instability is present at the level of the
conformal curvature tensor (Weyl tensor) - {\em cf.}\ the mass
inflation scenario inside charged spherical black holes
\cite{israel-poisson}. This is ruled out in Section 4 where we show
that all the gauge and tetrad invariant perturbed (Newman-Penrose)
Weyl scalars remain finite at the Cauchy horizon. Perturbations with
angular mode number $l=1$ require a separate (much shorter)
treament, which is carried out in Section 5 and yields the same
results. We make some concluding comments in Section 6. We note that
the analysis of the inhomogeneous wave equation below follows
closely our previous analysis of the minimally coupled massless
scalar wave equation in a general spherical self-similar spacetime
\cite{nolan-scalar}. We use the conventions of \cite{wald}, and set
$G=c=1$.
\section{Geometry of Vaidya Spacetime}
The Vaidya spacetime metric is the unique solution of the Einstein
equations subject to the assumptions that (a) spacetime is
spherically symmetric and (b) the energy-momentum tensor is that of
null dust, with the dust flow vector $\bar{k}^a$ normal to the
$SO(3)$ symmetry group orbits (the bar indicates a background
quantity). Local conservation of the energy momentum tensor shows
that $\bar{k}^a$ is geodesic and hypersurface orthogonal, and so one
can introduce a null coordinate $v$ with $\bar{k}_a=-\nabla_av$,
where we assume that $v$ increases into the future. The
hypersurfaces $v=$ constant are either future null cones of the axis
$\{r=0\}$, coresponding to expanding matter, or past null cones of
the axis, corresponding to collapsing matter. We assume the latter
(and so conform with the notation of the previous section). Taking
as the remaining coordinates the standard angular coordinates
$(\theta,\varphi)$ on the unit sphere and the areal radius $r$, the
line element can be shown to take the form \be ds^2 =
-(1-\frac{2m(v)}{r})dv^2+2dvdr+r^2d\Omega^2,\label{vaidyalel}\ee
where $d\Omega^2=d\theta^2+\sin^2\theta d\varphi^2$ is the standard
line element on the unit sphere. The energy-stress-momentum tensor
is obtained from $8\pi t_{ab}=\bar{\rho}\bar{k}_a\bar{k}_b,$ with
$\bar{\rho}=2m^\prime(v)/r^2$.  Then the strong, weak and dominant
energy conditions are all equivalent to $m'(v)\geq0$ for all $v$ in
the domain of $m$. The line element above is used to model the
gravitational collapse of a thick shell of null dust (or a photon
fluid) with the choice
\[ m(v) = \left\{ \begin{array}{cc}
                    0 & v<0; \\
                    M(v) & 0\leq v < v_1; \\
                    M(v_1) & v_1\leq v,
                  \end{array}
                  \right. \]
where $v\to M(v)$ is an increasing $C^1$ function on $[0,v_1)$ and
$v_1>0$ is arbitrary. Note then that spacetime is a portion of
Minkowski spacetime to the past of the past null cone $\{v=0\}$, and
is a portion of the Schwarzschild-Kruskal space-time with mass
parameter $M(v_1)$ to the future of the past null cone $\{v=v_1\}$.
The null fluid is confined to the region $\{0<v<v_1\}$, and
collapses from past null infinity to form a singularity at $r=0$.
The portion of the singularity in $v>0$ is future space-like, but
the singular origin $\{(v,r)=(0,0)\}$ may be visible (timelike or
ingoing null), depending on the details of the function $M$ at
$v=0$. See Figures 1 and 2\footnote{Note that the spacetime of
Figure 2 displays a somewhat curious feature of event horizons
related to their dependence on the global structure of spacetime:
the possibility of their appearing in a region of spacetime whose
causal past is flat.}.

\begin{figure}[h]
\centerline{\epsfxsize=6cm \epsfbox{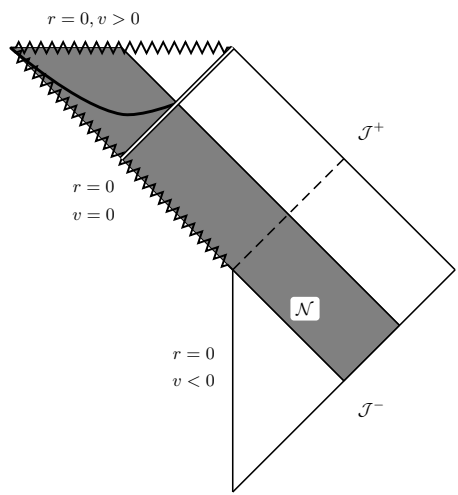}}
\caption{Conformal diagram for Vaidya-Schwarzschild collapse to a
globally naked singularity. We use the advanced Bondi co-ordinates
$v$ and $r$ described in the text. The Cauchy horizon is shown
dashed, the event horizon as a double line and the apparent horizon
as a bold curve. Note that the apparent horizon in the matter filled
region meets the event horizon at the surface of the `star'. $\pnc$
is the past null cone of the scaling origin. In Figures 1-3, the
matter filled region is shaded.}
\end{figure}

\begin{figure}[h]
\centerline{\epsfxsize=5cm \epsfbox{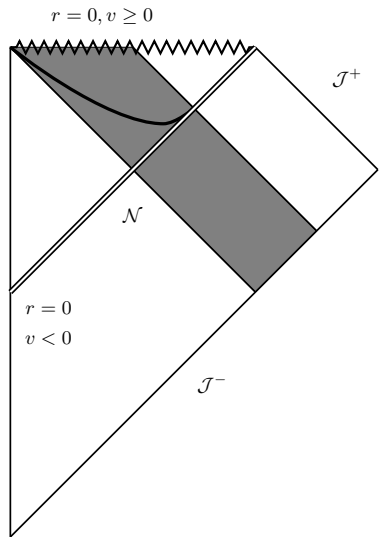}}
\caption{Conformal diagram for Vaidya-Schwarzschild collapse to a
black hole. As in Figure 1, the apparent horizon (bold) in the
matter filled region necessarily meets the event horizon (double
line). }
\end{figure}

In fact, we will not impose the cut off at $v=v_1$, and will take
the exterior region $v>0$ to be filled with null dust. That is, we
take
\[ m(v) = \left\{ \begin{array}{cc}
                    0 & v<0; \\
                    M(v) & 0\leq v.
                  \end{array}
                  \right. \]
We wish to study the stability of the Cauchy horizon in the case
when a naked singularity is present. In the cut-off spacetime, the
portion of the Cauchy horizon that resides in the
Schwarzschild-Kruskal region is (a portion of) a regular outgoing
null hypersurface of that spacetime where no singular behaviour can
be expected, unless divergence is mediated along the earlier portion
of the horizon that resides in the matter-filled region. Thus we are
only interested in the matter-filled region of the spacetime, and
the cut-off is unnecessary. So our concern is the spacetime of
Figure 3.

\begin{figure}[h]
\centerline{\epsfxsize=6cm \epsfbox{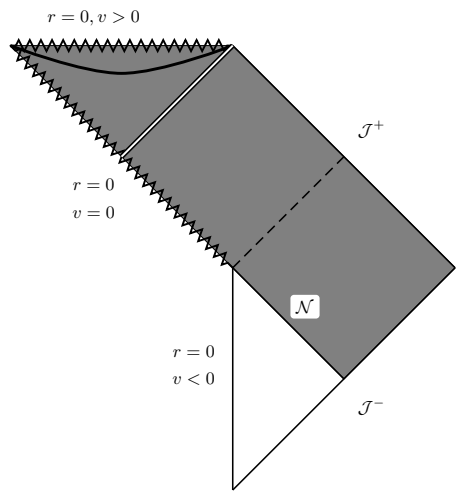}}
\caption{Conformal diagram for Vaidya collapse to a globally naked
singularity. The region to the future of $v=0$ is filled with null
dust.}
\end{figure}

This spacetime is self-similar when (and only when) $M$ is a linear
function of $v$: $M(v)=\lambda v$ for $\lambda>0$. The restriction
on the range of $\lambda$ ensures that the energy conditions are
satisfied, and that the trivial case is avoided ($\lambda=0$
corresponds to flat spacetime). Applying Proposition 2, it is then
straightforward to show that the singular origin is naked if and
only if $\lambda \in (0,\frac{1}{16})$, and we will assume
henceforth that $\lam$ lies in this range.

Introducing the coordinates $t=v/r, x=\log r$ of Section 1, the line
element (\ref{vaidyalel}) becomes \be ds^2 = e^{2x}[-(1-2\lam
t)dt^2+2(1-t+2\lam t^2)dtdx+t(2-t+2\lam t^2)dx^2+d\Omega^2].
\label{ssvaidya}\ee The Cauchy horizon is given by
\[ t=t_1:=\frac{1-\nu}{4\lam},\]
and the second future similarity horizon is given by \[
t=t_2:=\frac{1+\nu}{4\lam},\] where $\nu=\sqrt{1-16\lam}$. There are
no other future similarity horizons. The apparent horizon is
spacelike and is located at $t=t_3:=1/2\lam$. This forms to the
future of both future similarity horizons: $t_3>t_2>t_1$.

For the calculation of the gauge invariant perturbed Weyl curvature
scalars, it will be essential to have an appropriate representation
of the radial null directions of self-similar Vaidya spacetime,
along with the associated null coordinates. The advanced null
coordinate is $v=tr=te^x$, so that $v=$ constant describes a past
null cone of the axis $\{r=0\}$. The future null cones are described
by $u=$ constant, where the retarded null coordinate $u$ is given by
\be u = -e^x|t_1-t|^{\lam_1}|t_2-t|^{\lam_2},\label{udef}\ee with
\[ \lam_1=\frac{\nu+1}{2\nu},\quad
\lam_2=\frac{\nu-1}{2\nu}.\] The region of spacetime with which we
will be concerned is that bounded by past and future null infinity,
the past null cone of the scaling origin $\pnc=\ch^{-}$ and the
Cauchy horizon $\ch^+$. In the coordinates $(u,v)$, the
corresponding Lorentzian 2-space is
\[ {\cal{M}}_2=\{(u,v): -\infty<u<0, 0<v<+\infty.\}\] and we have
the following representations (see Figure 4):
\begin{eqnarray*}
{\cal{J}}^- & = & \{(u,v): u = -\infty, 0<v<+\infty\},\\
{\cal{J}}^+ & = & \{(u,v): -\infty<u<0, v=+\infty\},\\
{\cal{H}}^- & = & \{(u,v): -\infty<u<0, v=0\},\\
{\cal{H}}^+ & = & \{(u,v): u = 0, 0<v<+\infty\}.\end{eqnarray*}

\begin{figure}[h]
\centerline{\epsfxsize=6cm \epsfbox{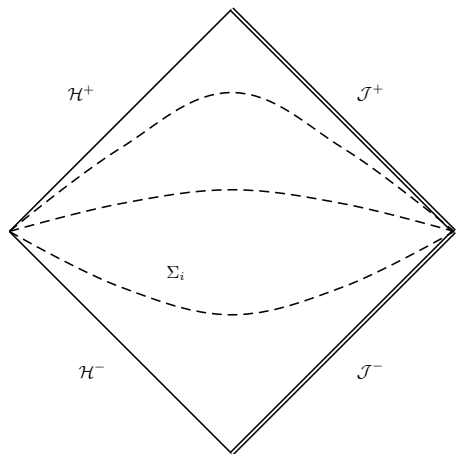}}
\caption{Conformal diagram of ${\cal{M}}_2$. Surfaces of constant
$t$ are shown dashed, including the initial data surface $\so$.
Surfaces at infinity are shown by double lines.}
\end{figure}

The radial null directions are $\frac{\partial}{\partial u}$ and
$\frac{\partial}{\partial v}$, and we choose the following scalings:
we will take $\vec{l}$ to be the future pointing ingoing radial null
vector field  given by \be \vec{l} =
-\frac{2}{H}\frac{\partial}{\partial u},\label{ldef}\ee where \be
H=H(t)=-2\lam|t_1-t|^{\lam_2}|t_2-t|^{\lam_1}\label{Hdef}\ee and we
take $\vec{n}$ to be the future pointing outgoing radial null vector
field given by \be \vec{n} =\frac{\partial}{\partial
v}.\label{ndef}\ee With these choices of scaling, $\vec{l}$ is an
affinely parametrized null geodesic tangent vector field, and the
Newman-Penrose normalization $g_{ab}l^an^b=-1$ holds. We note also
that in these coordinates, the line element is
\[ ds^2=Hdudv+r^2(u,v)d\Omega^2.\]

\section{Odd-Parity Perturbations}

\subsection{The Gerlach-Sengupta formalism}

To study perturbations of this spherically symmetric spacetime, we
will use the gauge invariant formalism introduced by Gerlach and
Sengupta \cite{GS}. (We follow the presentation of Martin-Garcia and
Gundlach \cite{jm+carsten2}.) This is based on the natural 2+2
splitting of a spherically symmetric spacetime, and a multipole
decomposition that enables an efficient treatment of the angular
dependence of the perturbation.

The metric of a spherically symmetric space-time $({\cal{M}}^4,g)$
can be written as \be
ds^2=g_{AB}(x^C)dx^Adx^B+r^2(x^C)\gamma_{\alpha\beta}dx^\alpha
dx^\beta, \label{bgmetric}\ee where $g_{AB}$ is a Lorentzian metric
on a 2-dimensional manifold with boundary $M^2$ and
$\gamma_{\alpha\beta}$ is the standard metric on the unit 2-sphere
$S^2$. Capital Latin indices represent tensor indices on $M^2$, and
Greek indices are tensor indices on $S^2$. $r(x^C)$ is a scalar
field on $M^2$. 4-dimensional space-time indices will be given in
lower case Latin. The covariant derivatives on $M^4$, $M^2$ and
$S^2$ will be denoted by a semi-colon, a vertical and a colon
respectively. $\epsilon_{AB}$ and $\epsilon_{\alpha\beta}$ are
covariantly constant anti-symmetric unit tensors with respect to
$g_{AB}$ and
$\gamma_{\alpha\beta}$. We define \bq v_A&=&\frac{r_{|A}}{r}, \label{vdef}\\
V_0&=&-\frac{1}{r^2}+2{v^A}_{|A}+3v^Av_A.\label{v0def}\eq Writing
the energy-momentum tensor as \be t_{ab}dx^a dx^b=
t_{AB}(x^C)dx^Adx^B+Q(x^C)r^2\gamma_{\alpha\beta}dx^\alpha
dx^\beta,\ee the Einstein equations of the spherically symmetric
background read \bq G_{AB}&=&-2(v_{A|B}+v_Av_B)+V_0g_{AB}=8\pi
t_{AB}\label{eeq1}\\ \frac12 G^\alpha_\alpha&=&
-{\cal{R}}+{v^A}_{|A}+v^Av_A=:8\pi Q,\label{eeq2}\eq where
$G^\alpha_\alpha=\gamma^{\alpha\beta}G_{\alpha\beta}$ and
${\cal{R}}$ is the Gaussian curvature of $M^2$.

Spherical symmetry of the background allows us to expand the
perturbed metric tensor in terms of spherical harmonics. Writing
$Y=Y^m_l$ and suppressing the indices $l,m$ throughout, we have the
following bases for scalar, vector and tensor harmonics
respectively: $\{Y\}$,
$\{Y_\alpha:=Y_{:\alpha},S_\alpha:=\epsilon_\alpha^\beta
Y_{\beta}\}$ and
$\{Y\gamma_{\alpha\beta},Z_{\alpha\beta}:=Y_{\alpha:\beta}+\frac{l(l+1)}{2}Y\gamma_{\alpha\beta},S_{\alpha:\beta}+S_{\beta:\alpha}\}$.
These are further classified depending on the transformation
properties under spatial inversion ${\vec x}\to -\vec{x}$ on the
unit sphere: a spherical harmonic with index $l$ is called even if
it transforms as $(-1)^l$ and is called odd if it transforms as
$(-1)^{l+1}$. In the bases above, $Y,Y_{\alpha}$ and
$Z_{\alpha\beta}$ are even and $S_\alpha,S_{(\alpha:\beta)}$ are
odd. We note that in the analysis below, the multipole index $l$
only appears in the combination $(l-1)(l+2)$ and so we define
${\ell}=(l-1)(l+2)$.

The perturbation $\delta g_{ab}$ of the metric tensor can then
be decomposed as \bq \delta g_{AB}&=&h_{AB}Y,\label{tensor}\\
\delta g_{A\beta}&=&h^E_AY_{:\beta}+h_A^OS_\beta,\label{vector}\\
\delta
g_{\alpha\beta}&=&r^2K\gamma_{\alpha\beta}Y+r^2GZ_{\alpha\beta}+2hS_{(\alpha:\beta)}.\label{scalar}
\eq The superscripts $E,O$ stand for even and odd respectively. Note
that $h_{AB}$, $\{h_A^E,h_A^O\}$ and $\{K,G,h\}$ are respectively a
2-tensor, vectors and scalars on $M^2$. A similar decomposition of
the perturbation of the stress-energy tensor is made:
\bq \delta t_{AB}&=&\Delta t_{AB}Y,\label{emt-tensor}\\
\delta t_{A\beta}&=&\Delta t^E_AY_{:\beta}+\Delta t_A^OS_\beta,\label{emt-vector}\\
\delta t_{\alpha\beta}&=&r^2\Delta t^3
\gamma_{\alpha\beta}Y+r^2\Delta t^2 Z_{\alpha\beta}+2\Delta
tS_{(\alpha:\beta)}.\label{emt-scalar} \eq In this case, $\Delta
t_{AB}$, $\{\Delta t_A^E,\Delta t_A^O\}$ and $\{\Delta t^3, \Delta
t^2, \Delta t\}$ are respectively a 2-tensor, vectors and scalars on
$M^2$.

A complete set of gauge invariant variables is produced as follows.
An infinitesmal co-ordinate transformation on the background is
generated by a vector field $\vec{\xi}$. Again, we can decompose
into even and odd harmonics and consider separately the
transformations generated by the 1-form fields \bq \bmt\xi^E
&=&\xi_A(x^C)Ydx^A+\xi^E(x^C)Y_{:\alpha}dx^\alpha,\label{evenxi}\\
\bmt\xi^O&=&\xi^OS_\alpha dx^\alpha. \label{oddxi}\eq From the
transformed versions of the metric perturbations, one can construct
combinations which are independent of the coefficients of
$\vec{\xi}$. These combinations are then gauge invariant. As we will
only be concerned with the odd parity sector, we give only these
terms. The entire odd parity metric perturbation is captured by the
gauge invariant co-vector field \bq
k_A&=&h_A^O-h_{|A}+2hv_A,\label{kvecdef} \eq and the gauge invariant
matter perturbation is described by \bq L_A &=&\Delta
t_A^O-Qh_A^O,\label{lvecdef}\\
L&=&\Delta t -Qh.\label{lscalardef}\eq

The linearized Einstein equations read \bq
\frac{1}{r^2}\left[r^4\left(\frac{k_A}{r^2}\right)_{|C}-r^4\left(\frac{k_C}{r^2}\right)_{|A}\right]^{|C}-\frac{{\ell}}{r^2}k_A=-16\pi
L_A,\label{veceqn}\\{k^A}_{|A}=16\pi L.\label{diveqn} \eq The latter
equation follows from the former and the linearized conservation
equation \be (r^2L_A)^{|A}={\ell}L.\label{lincons}\ee

As noted by Gerlach and Sengupta \cite{GS}, the vector equation
(\ref{veceqn}) is equivalent to the single scalar equation \be
\left[\frac{1}{r^2}(r^4\Pi)^{|A}\right]_{|A}-{\ell}\Pi=-16\pi\epsilon^{AB}L_{A|B},\label{master}\ee
where \be \Pi:=\epsilon^{AB}(r^{-2}k_A)_{|B} \label{pidef}\ee is a
gauge invariant scalar. $k_A$ is recovered from $\Pi$ and $L_A$
using \be {\ell}k_A=16\pi
r^2L_A-\epsilon_{AB}(r^4\Pi)^{|B}.\label{kvecredef}\ee The quantity
$\Pi$ is not only a gauge invariant scalar that, along with $L_A$,
completely determines the metric perturbation, but as shown in
\cite{nolan-weyl}, has the tetrad and gauge invariant interpretation
of being the perturbation of the Coulomb component $\Psi_2$ of the
background Weyl tensor.

In order to close the system of perturbation equations, an equation
of state must be given for the perturbed spacetime. With this
addition, the equations (\ref{lincons}) and (\ref{master})
completely determine the perturbation.

An important point to note is that the formalism described above is
incomplete for  $l=1$. (There is of course no odd parity $l=0$
perturbation.) For $l=1$, $h$ is not defined, being a coefficient of
zero, and so should be considered to be zero. Thus the gauge
invariants cannot be constructed. However it is convenient to use
the same variables (\ref{kvecdef})-(\ref{lscalardef}) for all values
of $l$. For $l=1$, these variables are only partially gauge
invariant, and so gauge-fixing is required. We defer treatment of
the case $l=1 (\ell=0)$ to Section 5, and so assume until then that
$l\geq2$.

\subsection{The matter perturbation}

We assume that the non-vacuum portion of the perturbed spacetime is
filled with null dust. This allows us to write
\[ t_{ab}= (\bar{\rho}+\delta\rho)(\bar{k}_a+\delta
k_a)(\bar{k}_b+\delta k_b),\] where the barred terms refer to
background quantities, so that $\bar{k}_a=-\nabla_av$. Retaining
only first order terms, we define
\[ \delta t_{ab}
=\delta\rho\bar{k}_a\bar{k}_b+\bar{\rho}(\bar{k}_a\delta
k_b+\bar{k}_b\delta k_a).\] Comparing with
(\ref{emt-tensor})-(\ref{emt-scalar}) and recalling that we are
considering only odd parity perturbations, we can determine the
gauge invariant matter perturbation using (\ref{lvecdef}) and
(\ref{lscalardef}): we find $L=0$ and \be
L_A=\mu(x^A)\bar{\rho}\bar{k}_A,\label{lvecform}\ee for some (first
order) scalar $\mu$. The evolution of $\mu$ is controlled by
(\ref{lincons}). In the self-similar coordinates $(t,x)$ we find \be
t\mu_{,t}-\mu_{,x} = 0,\label{mueqn}\ee which yields
\[ \mu = \mu_0(te^x) = \mu_0(v)\] for some arbitrary differentiable function $\mu_0$.
Thus the matter perturbation is completely determined by the
specification of the function $\mu_0$ on an initial slice of the
form $t=t_i\in(0,t_1)$.

\subsection{The master equation}

Having completely specified the matter perturbation in terms of an
initial data function, we now turn to the only remaining
perturbation equation (\ref{master}). We will refer to this as the
master equation and will work in the coordinates $(t,x)$ of Section
2. It is worth repeating that we deal only with the region $0<t<t_1$
on which $t$ is a time coordinate. The analysis below follows
closely that of \cite{nolan-scalar}, and where possible, we will
quote results from this paper rather than repeating very similar
proofs. We reiterate that $l\geq2$.

For fixed $\kp\in \mathbb{R}$, we define $\phi(t,x) = e^{\kp x}\Pi=
r^\kp\Pi$. Then, using (\ref{master}) and the line element
(\ref{ssvaidya}), we find that $\phi$ satisfies the inhomogeneous
wave equation \bq \alpha \phi_{,tt}+2\beta \phi_{,tx}+\gamma
\phi_{,xx} + (\alpha^\prime + (6-2\kp)\beta)\phi_{,t} +
(\beta^\prime +(6-2\kp)\gamma) \phi_{,x} \nonumber \\+
((4-\kp)\beta^\prime + (8-6\kp+\kp^2)\gamma-{\ell}e^\psi)\phi =
8\lam e^{(\kp-3)x}\mu_0(te^x),\label{main}\eq where $\psi=0$ and \bq
\alpha&=&-2t(1-tG),\label{aldef}\\
\beta&=&1-2tG,\label{bedef}\\
\gamma&=&2G,\label{gamdef}\\
G&=&\frac12-\lam t.\label{gdef} \eq Modulo the specification of the
functional form of $G$, the left hand side of (\ref{main}), along
with (\ref{aldef})-(\ref{gamdef}), gives the general form of the
left hand side of the master equation for the line element
(\ref{selfsimlel}).

We fix an initial data surface for (\ref{main}) given by
$\so=\{(t,x): t=t_i, x\in\mathbb{R}\}$ with $t_i\in(0,t_1)$. Our
principal concern is how $\phi$ and various of its derivatives
representing gauge invariant curvature scalars behave in the
approach to the Cauchy horizon, subject to initial regularity
conditions imposed at $\so$. Noting that we may write
\[ \alpha(t) = -2\lam t(t_1-t)(t_2-t)\] and so $\alpha(t_1)=0$,
we see that this question is rendered nontrivial by virtue of the
fact that the Cauchy horizon is a singular hypersurface for the
equation (\ref{main}): the spacelike surfaces $t=$constant become
characteristic (null) in the limit $t\to t_1$. Prior to the Cauchy
horizon, the evolution of $\phi$ proceeds smoothly:

\begin{theorem}\label{thm1}
Let $f,g,\mu_0\in C^\infty_0(\mathbb{R})$ and let $\kp\in
\mathbb{R}$. Then there exists a unique solution $\phi\in
C^\infty([t_i,t_1)\times \mathbb{R},\mathbb{R})$ of the initial
value problem consisting of the equation (\ref{main}) and the
initial data
\[ \phi|_\so = f,\quad \phi_{,t}|_\so = g.\]
Furthermore, the solution satisfies $\phi|_{t=t_0}\in
C^\infty_0(\mathbb{R})$ for all $t_0\in[t_i,t_1)$.\fin
\end{theorem}

The proof of this result is standard and is most easily obtained by
rewriting the equation (\ref{main}) as a first order symmetric
hyperbolic system for \be\vv = \left(
\begin{array}{c}
  \phi \\
  \alpha\phi_{,t}+\beta\phi_{,x} \\
  \phi_{,x}
\end{array}\right). \label{vvdef} \ee
See for example Chapter 12 of \cite{mcowen}. It is convenient to
rescale the time coordinate by defining \be \tau =
-\int_{t_i}^t\frac{ds}{\alpha(s)}.\label{taudef}\ee Then $\tau$ is
an analytic function of $t$ on $[t_i,t_1)$, $\tau(t_i)=0$ and
$\lim_{t\to t_1}\tau=+\infty$. The master equation (\ref{main}) can
be written in first order symmetric hyperbolic form
\[ \vv_{,\tau}+A\vv_{,x}+B\vv = \vec{\jmath} \]
where $A,B$ are smooth, bounded matrix functions of $\tau$ on
$[0,\infty)$ and $A$ is symmetric with real distinct eigenvalues.
The source term $\vec{\jmath}$ is given by
\[ \vec{\jmath}=\left( \begin{array}{c} 0\\ -8\lam\alpha(t)e^{(\kp-3)x}\mu_0(te^x)\\0\end{array}\right).\label{jdef}\]

We wish to analyse the behaviour of the solutions described by
Theorem 1, and so we assume until indicated otherwise that the
hypotheses of this theorem hold. Then $\vj\in
C^\infty_0([t_i,t_1)\times \mathbb{R},\mathbb{R}^3)$.

We define
\[ E_1(\tau)=E_1[\phi](\tau)=\int_\mathbb{R}\|\vv\|^2\,dx.\] The growth of
this energy norm is described by the following corollary, again a
standard result. We use the notation
$\|\vec{a}\|^2=(\vec{a},\vec{a})$ (Euclidean inner product) and $
\|\vec{a}\|_2^2 = \int_\mathbb{R}\|\vec{a}\|^2\,dx$ ($L^2$ norm).
The terms $C_0,C_1,...$ represent possibly different constants that
depend only on the metric function $G$ and the angular mode number
$l$.

\begin{corr}
$E_1[\phi](\tau)$ is differentiable on $[0,\infty)$ and satisfies
\be E_1[\phi](\tau)\leq
e^{B_0\tau}(E_1[\phi](0)+\int_0^\tau\|\vj(\sigma)\|_2^2\,d\sigma),\label{e1growth}\ee
where $B_0= \sup_{\tau>0}|I-2B| <+\infty$. Consequently,
\begin{eqnarray*}
\int_\mathbb{R}|\phi(t,x)|^2\,dx&\leq& e^{B_0\tau}(E_1[\phi](0)+\int_0^\tau\|\vj(\sigma)\|_2^2\,d\sigma),\\
\int_\mathbb{R}|\phi_{,x}(t,x)|^2\,dx&\leq&
e^{B_0\tau}(E_1[\phi](0)+\int_0^\tau\|\vj(\sigma)\|_2^2\,d\sigma),\\
\int_\mathbb{R}|\phi_{,t}(t,x)|^2\,dx&\leq&
C_1e^{C_0\tau}(E_1[\phi](0)+\int_0^\tau\|\vj(\sigma)\|_2^2\,d\sigma).
\end{eqnarray*}\fin
\end{corr}

The bounds on the $L^2$ norm of $\phi$ and its derivatives come
straight from the definition of $E_1(\tau)$: the third requires the
use of Minkowski's inequality and incorporates the exponential
growth of $\alpha^{-1}$ as $\tau\to\infty$. As in
\cite{nolan-scalar}, the growth of these norms in the approach to
the Cauchy horizon $\tau\to\infty$ is analysed using a second energy
integral.

Let \[D(t) =
{\ell}+(\kp-4)\beta^\prime(t)-(\kp^2-6\kp+8)\gamma(t),\] and for an
arbitrary positive, real-valued differentiable function $K(t)$
define \be E_2[\phi,\mu_0](t)=\int_\mathbb{R}
-\alpha\phi_{,t}^2+\gamma\phi_{,x}^2+D\phi^2+Ke^{2(\kp-3)x}\mu_0^2\,dx.\label{e2def}\ee

\begin{lemma}
Let $\kp\in[0,4]$. Then for all $t\in[t_i,t_1]$, $D(t)\geq 0$ and
$D^\prime(t)\leq 0$.
\end{lemma}

\noindent{\bf Proof:} From the definitions (\ref{bedef}) and
(\ref{gamdef}) we obtain
\[ D(t) = {\ell} - (\kp^2-5\kp+4)+2\lam t(\kp^2-4\kp).\]
Thus $D^\prime(t)=2\lam\kp(\kp-4)\leq0$. Then for $t<t_1$,
\begin{eqnarray*}
D(t) &\geq& D(t_1) = {\ell}-4+(5-8\lam t_1)\kp-(1-2\lam t_1)\kp^2\\
&\geq &(5-8\lam t_1)\kp-(1-2\lam t_1)\kp^2,
\end{eqnarray*}
where the second inequality uses $l\geq2$. This last expression,
considered as a quadratic function of $\kp$, is non-negative for
$\kp\in[0,\kp_*]$ where
\[ \kp_* = \frac{5-8\lam t_1}{1-2\lam t_1}=4 +
\frac{2}{1+\nu}>4.\] This yields $D(t)\geq0$ on the range indicated.
\fin

\begin{lemma}
Let $\kp\in[0,\kp_1)$ where
\[ \kp_1=\frac12(5-4\lam+\lam
t_1^2)=\frac{1+32\lam-32\lam^2-\nu}{16\lam}.\] Then there exists
$t_c=t_c(\kp)\in[t_i,t_1)$, a constant $C_0$ and a choice of $K$
such that $E_2[\phi,\mu_0](t)\geq0$ and $\frac{dE_2}{dt}\leq C_0E_2$
for all $t\in[t_c,t_1)$.
\end{lemma}

\noindent{\bf Proof:} Noting that $\kp_1<4$, we see that Lemma 1
applies and so non-negativity of $E_2$ is immediate. Write
$\tmu=e^{(\kp-3)x}\mu_0$. $E_2(t)$ is a smooth function of $t$, and
smoothness of the solution $\phi$ and of $\mu_0$ allow
differentiation under the integral sign. The resulting integral is
simplified in three steps: (i) integration by parts of the term
$\phi_{,x}\phi_{,xt}$ and the removal of a boundary term - permitted
as $\phi$ has compact support on each slice $t=$ constant; (ii)
removal of the term with $\phi_{,tt}$ by application of the equation
(\ref{main}); (iii) removal of a total derivative containing
$\phi_{,t}\phi_{,xt}$. This results in \begin{eqnarray*}
\frac{dE_2}{dt}&=&
\int_\mathbb{R}\left[(\alpha^\prime+2(6-2\kp))u_{,t}^2+2(\beta^\prime+(6-2\kp)\gamma)u_{,t}u_{,x}+\gamma^\prime
u_{,x}^2 +D^\prime u^2\right. \\ && \left. -16\lam\tmu
u_{,t}+2K\tmu\tmu_{,t}+K^\prime\tmu^2\right] \,dx.\end{eqnarray*} In
the next round of simplifications, we apply Lemma 1 $(D^\prime\leq
0)$, the Cauchy-Schwarz inequality in the form
\[ \int_\mathbb{R} 2\tmu u_{,t}\,dx \leq
\int_\mathbb{R}\tmu^2+u_{,t}^2\,dx\] and the equation satisfied by
$\tmu$ (obtained from (\ref{mueqn}))
\[ t\tmu_{,t}-\tmu_{,x}+(\kp-3)\tmu=0.\]
These yield \begin{eqnarray*} \frac{dE_2}{dt}
&\leq&\int_\mathbb{R}\left[(\alpha^\prime+2(6-2\kp)+8\lam)u_{,t}^2+2(\beta^\prime+(6-2\kp)\gamma)u_{,t}u_{,x}+\gamma^\prime
u_{,x}^2 \right. \\
&& \left. + (K^\prime -2(\kp-3)\frac{K}{t}+8\lam)\tmu^2\right] dx \\
&=:& \int_\mathbb{R} I\,dx.\end{eqnarray*}

For any constant $C>0$, define $I_R$ by $I=CI_{E_2}+I_R$, where
$I_{E_2}$ is defined so that $E_2=\int_\mathbb{R} I_{E_2}\,dx$. That
is, \begin{eqnarray*} I_R&=&
(\alpha^\prime+2(6-2\kp)+8\lam)u_{,t}^2+2(\beta^\prime+(6-2\kp)\gamma)u_{,t}u_{,x}+\gamma^\prime
u_{,x}^2 -CDu^2 \\
&& +(K^\prime-(C+\frac{2}{t}(\kp-3))K+8\lam)\tmu^2.\end{eqnarray*}
The Lemma is proven by showing that there is a choice $C_0>0$ of $C$
and a value $t_c\in[t_i,t_1)$  for which $I_R\leq0$ on $[t_c,t_1)$.

For any choice of $C$, there is a positive differentiable function
$K$ defined on $[t_i,t_1)$ for which the coefficient of $\tmu^2$ in
$I_R$ is negative for all $\kp$ in the range specified. Making this
choice and applying Lemma 1, we obtain

\begin{eqnarray*} I_R &\leq&
(\alpha^\prime+C\alpha+2(6-2\kp)+8\lam)u_{,t}^2+2(\beta^\prime+(6-2\kp)\gamma)u_{,t}u_{,x}+(\gamma^\prime-C\gamma)
u_{,x}^2 \\ &=:&
a(t)u_{,t}^2+b(t)u_{,t}u_{,x}+c(t)u_{,x}^2.\end{eqnarray*}

Consider next the quadratic form
\[ Q(X,Y;t) = a(t)X^2+b(t)XY+c(t)Y^2.\]
We find \by a(t_1)&=&2(4\kp-5+4\lambda+\lambda t_1^2),\\
b(t_1)&=&-\frac{4}{t_1}(2\kp-5-\lambda t_1^2),\\
c(t_1)&=&-2(\lambda+\frac{C}{t_1}). \ey We note that $c(t_1)<0$ for
any $C>0$ and $\lambda\in(0,1/16)$. The term $a(t_1)$ is negative
due to the assumed bound on $\kp$. Then the discriminant is also
negative if $C$ is chosen sufficiently large. So $Q(X,Y;t_1)$ is
negative definite. Then by continuity of the coefficients $a,b,c$,
the quadratic form $Q(X,Y;t)=a(t)X^2+b(t)XY+c(t)Y^2$ is negative
definite for all $t$ sufficiently close to $t_1$. That is, there
exists some $t_c<t_1$ such that $Q(X,Y;t)\leq 0$ for all $X,Y\in
\mathbb{R}$ and $t\in[t_c,t_1)$ with equality holding iff $X=Y=0$.
We note however that the value of $t_c$ will depend on $\kp$, with
$t_c\to t_1$ as $\kp\to \kp_1$. This however does not affect the
proof, which is now completed. \fin

\begin{remarklem}
{\em By minimising $\kp_1(\lam)$ for $\lam\in(0,1/16)$, we could
restate Lemma 2 with the simpler requirement $\kp\in[0,\frac52)$.}
\end{remarklem}

We can now give our first main result.

\begin{theorem}\label{thm2}
Let $\phi$ be a solution of (\ref{main}) that is subject to the
hypotheses of Theorem 1 and Lemma 2. Then the energy
$E_2[\phi,\mu_0](t)$ of the solution satisfies the {\em a priori}
bound
\[ E_2[\phi,\mu_0](t)\leq C_1  E_1[\phi](0) + C_2 J_{\kp}[\mu_0],\quad t\in[t_i,t_1)\]
where
\[ J_{\kp}[\mu_0]= \int_\mathbb{R} e^{2(\kp-3)x}\mu_0^2(t_ie^x)\,dx.\]
\end{theorem}

\noindent{\bf Proof:} We point out first how to convert the bounds
of Corollary 1 to {\em a priori} bounds. To do this, we exploit the
self-similar nature of the solution $\mu$ of the matter perturbation
equation (\ref{mueqn}). We have
\[ \|\vec{\jmath}\|^2_2(\tau) = 64\alpha^2(t)\int_\mathbb{R}
e^{2(\kp-3)x}\mu_0^2(te^x)\,dx.\] A change of variable in the
integral yields
\[\|\vec{\jmath}\|^2_2(\tau) =
\left(\frac{\alpha(t)}{\alpha(t_i)}\right)^2\left(\frac{t_i}{t}\right)^{2(\kp-3)}J_{\kp}[\mu_0].\]
Then making the appropriate change of variables using
(\ref{taudef}), we obtain \begin{eqnarray*}
\int_0^\tau\|\vec{\jmath}\|^2_2(\sigma)\,d\sigma &=&J_{\kp}[\mu_0]
\int_{t_i}^{t}\frac{-\alpha(s)}{(\alpha(t_i))^2}\left(\frac{t_i}{s}\right)^{2(\kp-3)}\,ds
\\
&=:&h(t)J_{\kp}[\mu_0]\leq C_0 J_{\kp}[\mu_0]. \end{eqnarray*} From
Corollary 1, we can write \be E_2[\phi,\mu_0](t)\leq
d(t)(E_1[\phi](0)+C_2J_{\kp}[\mu_0]),\label{e2grow1}\ee where $d(t)$
is a smooth positive function of $t$ that diverges in the limit
$t\to t_1$. However, $d(t_c)$ is finite, where $t_c$ is the value of
$t$ identified in Lemma 2. From time $t_c$ onwards, $E_2$ obeys the
differential inequality of this lemma, which may be integrated to
yield \be E_2(t)\leq e^{C_0(t-t_c)}E_2(t_c),\quad
t\in[t_c,t_1).\label{e2grow2}\ee Combining (\ref{e2grow1}) and
(\ref{e2grow2}) yields the desired result. \fin

\begin{theorem}\label{thm3}
Let $\phi$ be a solution of (\ref{main}) that is subject to the
hypotheses of Theorem 1 and Lemma 2. Then $\phi$ is uniformly
bounded on $[t_i,t_1)\times \mathbb{R}$: there exist positive
constants $C_1$ and $C_2$ such that \be |\phi(t,x)|\leq
C_1E_1[\phi](0)+C_2J_{\kp}[\mu_0],\quad t\in[t_i,t_1),
x\in\mathbb{R}.\label{pointbnd}\ee
\end{theorem}

\noindent{\bf Proof:} Theorem 2 provides an {\em a priori} bound for
the $H^{1,2}$ norm of $\phi$: for all $t\in[t_i,t_1)$,
\[ \int_\mathbb{R} \phi^2 + \phi_{,x}^2 dx \leq C_1E_1[\phi](0) +
C_2J_{\kp}[\mu_0].\] The pointwise bound arises immediately on
application of the Sobolev inequality for $v\in
C^\infty_0(\mathbb{R})$:
\[ |v(x)|^2\leq\frac{1}{2}\left\{
\int_\mathbb{R}|v(y)|^2+|v^\prime(y)|^2\,dy\right\}.\] See p.\ 1057
of \cite{wald2} for a proof of this inequality. \fin

\begin{remarkthe}
{\em Note that by differentiating (\ref{main}) with respect to $x$,
we can obtain results similar to Theorems 1-3 for any spatial
derivative of $\phi$. The bounding term in the inequalities
corresponding to (\ref{pointbnd}) will involve sums of terms of the
form $E_1[\frac{\partial^n\phi}{\partial x^n}](0)$ and
$J_{\kp}[\mu_0^{(n)}]$.}\end{remarkthe}

\begin{remarkthe}
{\em Of principal concern in this paper is the behaviour of the
field $\Pi$ and those of its derivatives representing the perturbed
Weyl curvature scalars. Theorem 3 shows that $\Pi$ is bounded in the
limit as the Cauchy horizon is approached ($t\to t_1$). However this
does not imply that $\lim_{t\to t_1}\Pi(t,x)$ exists for any $x\in
\mathbb{R}$. We will show now that this is in fact the case; indeed
we can show that $\Pi|_{t=t_1}\in C^\infty(\mathbb{R})$. In
\cite{nolan-scalar}, the corresponding limit function was
erroneously assumed to exist on the basis of the equivalent to
Theorem 3. This assumption can be shown to be true by applying the
argument below to that paper, and so does not affect the results of
that paper.}
\end{remarkthe}

In order to get from the bound of Theorem 3 to the existence of the
limit $\lim_{t\to t_1}\phi(t,x)$, we need some control over the time
derivative of $\phi$ as the Cauchy horizon is approached. This is
provided by the following lemma, which relies on treating
(\ref{main}) as a first order transport equation for $\phi_{,t}$.

\begin{lemma}
Let $\phi$ be a solution of (\ref{main}) that is subject to the
hypotheses of Theorem 1 and Lemma 2. Then $\phi_{,t}$ is uniformly
bounded on $[t_i,t_1)\times \mathbb{R}$: there exist positive
constants $C_0,C_1,...$ such that \by |\phi_{,t}(t,x)|&\leq&
C_0E_1[\phi](0)+C_1E_1[\phi_{,x}](0)+C_2E_1[\phi_{,xx}](0)\\&&+C_3J_{\kp}[\mu_0]+C_4J_{\kp}[\mu_0^\prime]+C_5J_{\kp}[\mu_0^{\prime\prime}],\quad
t\in[t_i,t_1), x\in\mathbb{R}.\ey
\end{lemma}

\noindent{\bf Proof:} Define $\chi=\phi_{,t}$. Then (\ref{main}) can
be written \be \alpha\chi_{,t}+2\beta\chi_{,x}+(\alpha^\prime +
(6-2\kp)\beta)\chi=f(t,x),\label{trans}\ee where the right hand side
depends linearly on $\mu_0$ and the zeroth, first and second spatial
derivatives of $\phi$. Then $f$ is smooth and has compact support on
each slice $t=$constant. If we write (\ref{main}) as \[
L[\phi]=a_0(t,x)\mu_0(te^x),\] then differentiation with respect to
$x$ shows that $\phi_{,x}$ and $\phi_{,xx}$ satisfy equations with
identical first and second order derivative coefficients:
\by L[\phi_{,x}]&=&b_0(t,x)\mu_0(te^x)+b_1(t,x)\mu_0^\prime(te^x),\\
L[\phi_{,xx}]&=&c_0(t,x)\mu_0(te^x)+c_1(t,x)\mu_0^\prime(te^x)+c_2(t,x)\mu_0^{\prime\prime}(te^x).\ey
We can therefore apply Theorem 3 to $\phi_{,x}$ and $\phi_{,xx}$:
the only difference in the result will be that the bounding terms
will depend also on the $L^2$ norms of the first and second
derivatives of $\phi$ and $\mu_0$ at $t=t_i$. Then by linearity, we
can bound $f$ by a similar {\em a priori} term. The bound for $u$
then arises by straightforward integration of the first order
transport equation (\ref{trans}). See Theorem 6 of
\cite{nolan-scalar}. \fin

\begin{theorem}\label{thm4}
Let $\phi$ be a solution of (\ref{main}) that is subject to the
hypotheses of Theorem 1 and Lemma 2. Then $\phi_\chp:=\lim_{t\to
t_1}\phi(t,\cdot)\in C^\infty(\mathbb{R})$ and satisfies the bound
\[ |\phi_\chp(x)|\leq C_1E_1[\phi](0)+C_2J_{\kp}[\mu_0],\quad
x\in\mathbb{R}.\]
\end{theorem}

\noindent{\bf Proof:} Fix $x\in\mathbb{R}$ and consider a sequence
$\{t^{(n)}\}_{n=1}^\infty\subset[t_i,t_1)$ that converges to $t_1$.
For all $m,n\geq 1$, we can apply the mean value theorem to get \be
|\phi(t^{(m)},x)-\phi(t^{(n)},x)| =
|\phi_{,t}(t_*,x)||t^{(m)}-t^{(n)}|\label{cauchy}\ee for some $t_*$
between $t^{(m)}$ and $t^{(n)}$. Using the bound of Lemma 3, we see
that $\phi(t^{(m)},x)$ is a Cauchy sequence of real numbers, and so
for each $x\in\mathbb{R}$, $\lim_{t\to t_1}\phi(t,x)$ exists. Hence
$\phi_\chp$ is defined. We can apply an analogous argument to all
the spatial derivatives of $\phi$. It remains to show that
\[ \frac{d}{dx}\left\{\phi_\chp\right\} = \lim_{t\to
t_1}\phi_{,x}(t,x).\] (Again, an analogous argument will apply to
higher spatial derivatives.) But this follows by uniform convergence
of the sequence of functions $\phi(t^{(n)},x)$ to $\phi_\chp$, which
in turn follows from (\ref{cauchy}) and the uniform bound of Lemma
3. To obtain the bound in the statement of the theorem, we take the
limit $t\to t_1$ of the corresponding bound in Theorem 3. This is
permitted as the bounding term is independent of $t$. \fin

We conclude this section by extending the results of Theorems 1-4 to
the case where the initial data lie in appropriate Sobolev spaces.
This is important as it will allow the perturbation to be non-zero
at the axis $r=0$. The results so far relate to the case where the
initial data and the corresponding solutions are supported away from
$r=0$. This is an undesirable feature, as we would ideally like to
consider a perturbation that arises from data imposed on a globally
regular initial data slice of the space-time: our $t=t_i$ slice is
singular at the origin. Such a regular slice would intersect the
past null cone, and we should certainly consider the case where the
support of the initial data also does so. We can deal with such data
by taking the limit of a sequence of test function ($C^\infty_0$)
data in an appropriate Sobolev space.

\begin{theorem}\label{thm5}
Let $\kp\in[0,\kp_1)$ and let $\phi=e^{\kp x}\Pi$ and define
$\tmu(x)=e^{(\kp-3)x}\mu_0(x)=e^{(\kp-3)x}\mu(t_ie^x)$.
\begin{itemize}
\item[(i)] Let $f\in H^{1,2}(\mathbb{R})$, $g\in L^2(\mathbb{R})$,
$\tmu\in L^2(\mathbb{R})$. Then there exists a unique solution
$\phi\in C([t_i,t_1),H^{1,2}(\mathbb{R}))$ of the initial value
problem consisting of (\ref{main}) with initial data
$\phi|_{\so}=f$, $\phi_{,t}|_\so=g$. The solution satisfies the {\em
a priori} bound
\[ |\phi(t,x)|\leq C_0E_1[\phi](0)+C_2J_\kp[\tmu],\quad
t\in[t_i,t_1), x\in\mathbb{R}.\]
\item[(ii)] Let $f\in H^{3,2}(\mathbb{R})$, $g\in
H^{2,2}(\mathbb{R})$, $\tmu\in H^{2,2}(\mathbb{R})$. Then there
exists a unique solution $\phi\in C([t_i,t_1],H^{1,2}(\mathbb{R}))$
of the initial value problem consisting of (\ref{main}) with initial
data $\phi|_{\so}=f$, $\phi_{,t}|_\so=g$. The solution satisfies the
{\em a priori} bound
\[ |\phi(t,x)|\leq C_0E_1[\phi](0)+C_2J_\kp[\tmu],\quad
t\in[t_i,t_1], x\in\mathbb{R},\] and its time derivative satisfies
\by |\phi_{,t}(t,x)|&\leq&
C_0E_1[\phi](0)+C_1E_1[\phi_{,x}](0)+C_2E_1[\phi_{,xx}](0)\\&&+C_3J_{\kp}[\tmu]+C_4J_{\kp}[\tmu^\prime]+C_5J_{\kp}[\tmu^{\prime\prime}],\quad
t\in[t_i,t_1], x\in\mathbb{R}.\ey
\end{itemize}
\end{theorem}

\noindent{\bf Proof:} We give just the outline of the proof, which
is nearly identical to Theorems 5 and 7 of \cite{nolan-scalar} and
which relies on a standard PDE technique. For part (i), we consider
sequences of test functions $\{f_{(n)}\}_{n=0}^\infty$,
$\{g_{(n)}\}_{n=0}^\infty$, $\{{\tmu}_{(n)}\}_{n=0}^\infty$ and
apply Theorems 1-3 to obtain a sequence of smooth solutions
$\{\phi_{(n)}\}_{n=0}^\infty\subset
C^\infty([t_i,t_1)\times\mathbb{R})$ satisfying the bounds of
Theorems 2 and 3 above. By exploiting linearity of the equation and
the fact that $C^\infty_0(\mathbb{R})$ is dense in the Banach spaces
$L^2$ and $H^{k,2}$ for $k=1,2,3\dots$, we can legitimately take the
limit of relevant inequalities to prove the stated results. The
proof of part (ii) is similar, but higher order Sobolev spaces must
be invoked due to the form of the inequality of Lemma 2.\fin

\section{Gauge Invariant Curvature Scalars}
As seen above, the odd parity perturbation of Vaidya spacetime is
completely described by the vector $L_A$ and the scalar $\Pi$. The
latter quantity plays a dual role: on the one hand it is a potential
for the gauge invariant metric perturbation $k_A$ (see
(\ref{kvecredef})) and on the other, is the gauge invariant
perturbation of the Coulomb component $\Psi_2$ of the Weyl tensor
\cite{nolan-weyl}. For at least two reasons, it is desirable to have
a full set of gauge invariant scalars that describes the
perturbation of the Weyl tensor.

First, one prefers scalars as these avoid the problems presented by
an inappropriate choice of coordinates. The components of a non-zero
rank tensor may blow-up, with the blow-up inadvertently ascribed to
singular behaviour rather than the wrong choice of coordinates.

Second, the metric and matter perturbations alone do not capture the
whole physical picture (neither of course does the Weyl tensor
alone). Perhaps the best example of this is the case of
perturbations impinging on the inner (Cauchy) horizon of a charged
or spinning black hole. Here the metric perturbation remains
continuous (but not differentiable) and the Weyl curvature blows up,
a scenario described as mass inflation \cite{israel-poisson}. This
has been described rigourously in \cite{dafermos-mi}.

As shown in \cite{nolan-weyl}, the perturbed Weyl scalars can be
defined in a gauge invariant manner in the case of odd parity
perturbations. Using a null tetrad
$\{\vec{l},\vec{n},\vec{m},\vec{m}^*\}$ where the asterisk
represents complex conjugation and following the notation of
\cite{stewart} for the Weyl scalars $\Psi_0,\dots,\Psi_4$, the
perturbed Weyl scalars are given by \bq
\delta\Psi_0&=&\left.\frac{Q_0}{2r^2}\right.l^Al^Bk_{A|B},\label{oddp0}\\
\delta\Psi_1&=&\left.\frac{Q_1}{r}\right.\left[(r^2\Pi)_{|A}l^A-\frac{4}{r^2}k_Al^A\right],\label{oddp1}\\
\delta\Psi_2&=&{Q_2}\Pi,\label{oddp2}\\
\delta\Psi_3&=&\left.\frac{Q^*_1}{r}\right.\left[(r^2\Pi)_{|A}n^A-\frac{4}{r^2}k_An^A\right],\label{oddp3}\\
\delta\Psi_4&=&\left.\frac{Q^*_0}{2r^2}\right.n^An^Bk_{A|B},\label{oddp4}
\eq where the functions $Q_i, i=0,1,2$ depend only on the angular
coordinates. In this definition, we restrict to the preferred sets
of null tetrads defined on the spherically symmetric background for
which $\vec{l}$ and $\vec{n}$ are the principal null directions of
the background. Then there remains a scaling freedom in these
definitions. Under the spin-boost transformation
\[ (\vec{l},\vec{n}) \to (A\vec{l},A^{-1}\vec{n})\]
we find \[ \delta\Psi_i \to A^{2-i}\delta\Psi_i\] for $i=0,\dots 4$.
Thus we cannot ascribe direct physical significance to the values of
$\delta\Psi_i$ (except for $i=2$). However the terms \bq \delta
P_{-1} &=& |\delta\Psi_0\delta\Psi_4|^{1/2},\\ \delta P_0 &=&
\delta\Psi_2,\\ \delta P_1&=&|\delta\Psi_1\delta\Psi_3|^{1/2} \eq
are {\em fully} invariant perturbation scalars: i.e.\ they are first
order scalars which are both gauge and tetrad invariant.

\subsection{The master equation in null coordinates}

It is straightforward to calculate $\delta\Psi_0,\dots\delta\Psi_4$
in the coordinates $(t,x)$ of Section 3. However it is less
straightforward to determine whether or not the resulting quantities
are finite at the Cauchy horizon $t=t_1$: we encounter terms
involving products of negative powers of the term $H(t)$ of Section
2 with first and second time $(t)$ derivatives of $\Pi$. While we
know that the first time derivative of $\Pi$ is bounded, we have no
information regarding the behaviour of the second time derivative at
the Cauchy horizon. It turns out that we can circumvent this problem
by calculating the scalars (\ref{oddp0})-(\ref{oddp4}) in the null
coordinates $(u,v)$ of Section 2. This approach also necessitates
rewriting the equation (\ref{main}) in these null coordinates. The
result of this is as follows. Let
\[ \Psi = r^3\Pi = e^{3x}\Pi.\]
Then we find \be
\Psi_{,uv}+\tilde{V}(u,v)\Psi=\tilde{F},\label{weqnull1}\ee where
\by \tilde{V} &=&
-\frac{1}{4uv}(\alpha\beta^\prime+1-\beta^2-{\ell}\alpha),\\
\tilde{F}&=&-2\lam\frac{\alpha}{uv}\mu_0(v).\ey In these
coordinates, we find that \be
\delta\Psi_0=-\frac{4}{H^2}(\Psi_{,uu}+\frac{2-\alpha^\prime}{2u}\Psi_{,u})Q_1\label{oddp0null1}\ee
for which the problem mentioned above remains. However the following
relabelling of the null cones resolves the problem.

Define \be
X=|u|^{\frac{1}{\lam_1}}=|t_1-t||t_2-t|^{\frac{\nu-1}{\nu+1}}\exp(\frac{x}{\lam_1}),\quad
Y=v=te^x.\label{bestnull}\ee We note that $X=0$ at the Cauchy
horizon $t=t_1$. Then (\ref{weqnull1}) - that is, the master
equation - takes the form \be \Psi_{,XY}+V(X,Y)\Psi =
F(X,Y),\label{weqnull2}\ee where
\[ V(X,Y) =
-\frac{1}{2}\lam_1\lam|t_2-t|^{\frac{2}{1+\nu}}\exp(-(\frac{1+3\nu}{1+\nu})x)\]
and
\[ F(X,Y) =4\lam_1\lam^2|t_2-t|^{\frac{2}{1+\nu}}\exp(-(\frac{1+3\nu}{1+\nu})x)\mu_0(Y).\]

Note that for any fixed $Y=v>0$, $V$ and $F$ are analytic functions
of $X$ at $X=0$. Moreover, there exists $C>0$ and sequences of
smooth functions $\{V_n\}_{n=0}^\infty$, $\{F_n\}_{n=0}^\infty$ such
that \be V(X,Y) = \sum_{n=0}^\infty V_n(Y)X^n,\quad F(X,Y) =
\sum_{n=0}^\infty F_n(Y)X^n,\quad|X|<C. \label{vfexp}\ee Combining
the high degree of regularity of the coefficients of
(\ref{weqnull2}) with the results of Section 3 yields the following.

\begin{theorem}\label{thm6}
Let $f,g,\mu_0\in C^\infty_0(\mathbb{R})$ and let $\Pi\in
C^\infty([t_i,t_1)\times \mathbb{R})\cap C^\infty(\{t=t_1\}\times
\mathbb{R})$ be the unique solution of the initial value problem
consisting of (\ref{main}) with initial data
\[ \Pi|_\so = f,\quad \Pi_{,t}|_\so = g.\] Let $X,Y$ be as defined
in (\ref{bestnull}). Then there exists $X_0>0$ such that
$\Psi=e^{3x}\Pi$ satisfies $\Psi\in C^\infty(\Omega)$ where
\[ \Omega =\{(X,Y): 0\leq X\leq X_0, Y>0\}.\]
Thus all $X\!\!-$ and $Y\!\!-$derivatives of $\Psi$ are finite at
the Cauchy horizon $X=0$.
\end{theorem}

\noindent{\bf Proof:} Consider the characteristic rectangle \[
R=\{(X,Y): 0\leq X\leq X_0, Y_0\leq Y\leq Y_1\}\] where $X_0>0$ and
$0<Y_0<Y_1$. Applying Theorem 1 and Theorem 4 with $\kp=0$ and
noting that the coordinate transformation $(t,x)\to(u,v)$ is a
homeomorphism on $R$ for sufficiently small $X_0$, we see that $\Psi
\in C^0(R,\mathbb{R})$. Then rewriting (\ref{weqnull2}) as
\[ \Psi_{,XY}=-V\Psi+F=:Q(X,Y),\]
we have $Q\in C^0(R,\mathbb{R})$. For any $Y\in(Y_0,Y_1]$, we can
then integrate to obtain
\[ \frac{\partial \Psi}{\partial X}(X,Y)-\frac{\partial \Psi}{\partial
X}(X,Y_0)=\int_{Y_0}^Y Q(X,Z)\,dZ.\] We can choose $Y_0$ to be small
enough so that the ingoing null ray $Y=Y_0$ lies outside the support
of $\Psi$ (see Figure 5). Then $\frac{\partial \Psi}{\partial
X}(X,Y_0)=0$, and
\[ \frac{\partial \Psi}{\partial X}(X,Y)=\int_{Y_0}^Y Q(X,Z)\,dZ,\] giving $\Psi_{,X} \in
C^0(R,\mathbb{R})$. A similar argument yields $\Psi_{,Y}\in
C^0(R,\mathbb{R})$, and so $\Psi\in C^1(R,\mathbb{R})$. Continuing
this argument inductively, we obtain $\Psi\in
C^\infty(R,\mathbb{R})$ for all $Y_0$ taken sufficiently small and
all $Y_1>0$. \fin

\begin{figure}[h]
\centerline{\epsfxsize=10cm \epsfbox{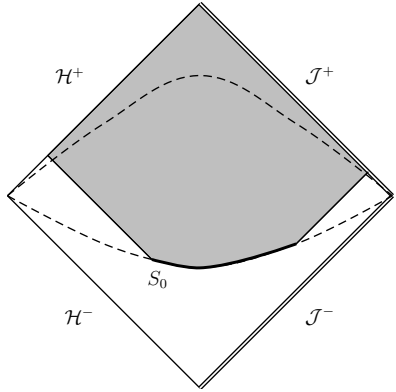}}
\caption{Spread of the support of the field (shaded), with initial
data with compact support $S_0\subset \Sigma_i$. There are ingoing
null rays from $\Sigma_i$ to ${\cal{H}}^+$ which lie outside the
support of the field.}
\end{figure}

\begin{remarkthe}{\em It is reasonable to ask why we have not simply
stated the entire problem in the coordinates $(X,Y)$ and deduced
finiteness of $\Psi$ - and hence $\Pi$ - at the Cauchy horizon by
writing down a very simple existence and uniqueness theorem for the
characteristic initial value problem consisting of (\ref{weqnull2})
with characteristic data $\Psi|_{X=X_0}, \Psi|_{Y=Y_0} \in
C^\infty(0,+\infty)$. The answer to this is that this formulation of
the problem assumes that the field $\Pi$ is a smooth function of the
retarded null coordinate $X$ {\em at} the Cauchy horizon. But the
question of whether or not this is a valid assumption is the very
point that we are attempting to address here, with respect to finite
initial Cauchy data posed on a hypersurface that precedes the Cauchy
horizon.}
\end{remarkthe}

\subsection{The perturbed Weyl scalars} There is a further
advantage of the coordinates $(X,Y)$. Not only does the master
equation assume the simple form (\ref{weqnull2}), but we find that
the perturbed Weyl scalars assume a very simple form when expressed
in these coordinates. Using (\ref{oddp0null1}) and (\ref{bestnull}),
we find \be \delta\Psi_0 =
-\frac{4}{\lam_1^2H^2}X^{\frac{\nu-1}{\nu}}(\Psi_{,XX}+(1-\frac{\lam_1}{2}\alpha^\prime)X^{-1}\Psi_{,X})e^{-x}Q_0.\label{oddp0null2}\ee
Using (\ref{Hdef}), (\ref{aldef}), and (\ref{bestnull}), we have \by
H^{-2}X^{\frac{\nu-1}{\nu}}&=&\frac{1}{4\lam^2}|t_2-t|^{-\frac{4}{1+\nu}}\exp(2(\frac{\nu-1}{\nu+1})x),\\
(1-\frac{\lam_1}{2}\alpha^\prime)X^{-1}&=&\lam_1\lam(2t_2-t_1-3t)(t_2-t)^{\frac{\nu-1}{\nu+1}}\exp(-(\frac{2\nu}{1+\nu})x).\ey
Thus $\delta\Psi_0$ is finite at the Cauchy horizon $X=0 (t=t_1)$.
This also holds for the other perturbed Weyl curvature scalars.
Finiteness is immediate for $\delta\Psi_2$, which is essentially
$\Pi$. The other gravitational radiation scalar, representing
outgoing radiation, is found to be \be \delta\Psi_4 =
\{(\Psi_{,YY}+(1-4\lam t)e^{-x}\Psi_{,Y}-4\lam
e^{-2x}\Psi)e^x-4\lam(\mu_0^\prime(Y)-2\lam
te^{-x}\mu_0(Y))\}e^{-2x}Q_0^*.\label{oddp4null2}\ee It is immediate
from Theorem 6 that this is finite at the Cauchy horizon. For the
two remaining scalars, we find \be \delta\Psi_1 = \left\{
\frac{2}{\lam_1}({\ell}-4)H^{-1}X^{\frac{\nu-1}{2\nu}}e^x\Psi_{,X}+({\ell}+4)\Psi\right\}
e^{-3x}\frac{Q_1}{{\ell}}\label{oddp1null2}\ee with
\[
H^{-1}X^{\frac{\nu-1}{2\nu}}=-\frac{1}{2\lam}|t_2-t|^{-\frac{2}{\nu+1}}\exp((\frac{\nu-1}{\nu+1})x),\]
while \be \delta\Psi_3=\left\{
({\ell}+4)\Psi_{,Y}-\frac{1}{2}({\ell}-4)t(1-2\lam
t)Y^{-1}\Psi+64\pi\bar{\rho}e^x\mu_0(Y)\right\}e^{-2x}\frac{Q_1^*}{{\ell}}.\label{oddp3null2}\ee
Thus these scalars are also finite at the Cauchy horizon. Thus we
have proven the following.

\begin{theorem}\label{thm7}
Subject to the hypotheses of Theorem 6, the perturbed Weyl scalars
(\ref{oddp1})-(\ref{oddp4}), calculated with respect to the null
vectors $\{\vec{l},\vec{n}\}$ defined in (\ref{ldef}) and
(\ref{ndef}), are finite at all points of $\{(t,x):t_i\leq t\leq
t_1,x\in\mathbb{R}\}$. In particular, the perturbed Weyl scalars are
finite at the Cauchy horizon $\chp$.\fin
\end{theorem}

\begin{remarkthe}{\em Having found a null tetrad in which {\em all}
the perturbed Weyl scalars are finite at the Cauchy horizon, it is
clear that the fully invariant scalars $\delta P_i, i=0,\pm1$ are
also finite thereat. A difficulty of interpretation of the
$\delta\Psi_i$ would only arise if one found (say) vanishing of
$\delta\Psi_0$ and divergence of $\delta\Psi_4$ at the Cauchy
horizon. In such a case, recourse to the calculation of $\delta
P_{-1}$ would be essential.}
\end{remarkthe}

As in the previous section, we wish to extend the present results to
initial data of the type considered in Theorem 5. As in the proof of
that theorem, the principal requirement is the existence of {\em a
priori} bounds for the relevant quantities.

\begin{theorem}\label{thm8}
Subject to the hypotheses of Theorem 6, the
following {\em a priori} bounds hold: for each $i\in\{0,\dots,4\}$,
there exist constants $C_0^{(i)},C_1^{(i)},\dots,C_5^{(i)}$ that
depend only on $\lam$ and $\kp$ such that for all
$(t,x)\in[t_i,t_1]\times\mathbb{R}$, \bq |e^{\kp
x}\delta\Psi_i(t,x)|&\leq&
C_0^{(i)}E_1[\phi](0)+C_1^{(i)}E_1[\phi_{,x}](0)+C_2^{(i)}E_1[\phi_{,xx}](0)\nonumber\\&&+C_3^{(i)}J_{\kp}[\mu_0]+C_4^{(i)}J_{\kp}[\mu_0^\prime]+C_5^{(i)}J_{\kp}[\mu_0^{\prime\prime}].
\label{apbound}\eq
\end{theorem}

\noindent{\bf Proof:} We consider first the case $\kp=0$. For $i=2$,
the bound (\ref{apbound}) is immediate from the definition
(\ref{oddp2}) and from Theorems 2 and 3 (where we let $\kp=0$ in
those theorems). A straightforward calculation shows that
\[ \delta\Psi_1= \left[
a(t)\Pi_{,t}+b(t)\Pi_{,x}+c(t)\Pi\right]\frac{Q_1}{{\ell}},\] where
here and in the rest of this proof, $a,b,c,\dots$ represent
functions of $t$ that are smooth and uniformly bounded on
$[t_i,t_1]$, and which may change from line to line. Similarly, we
find
\[ \delta\Psi_3= \left[
a(t)\Pi_{,t}+b(t)\Pi_{,x}+c(t)\Pi+128\pi\lam
e^{-3x}\mu_0(te^x)\right]\frac{Q_1}{{\ell}}.\] The result follows by
application of Theorems 2 and 3 to $\Pi$ and analogous results for
$\Pi_{,x}$ (see Remark 3.1) and by application of Lemma 3 which
provides the bound for $\Pi_{,t}$. Again, we take $\kp=0$ in these
results.

To obtain bounds on $\delta\Psi_0$ and $\delta\Psi_4$, we exploit
the form (\ref{weqnull2}) that the master equation takes in the null
coordinates $(X,Y)$. Integrating, and using the fact that $\Psi$ is
identically zero for sufficiently small values of $Y$, we can write
\[ \Psi_{,X}(X,Y)=\int_0^Y - V(X,Z)\Psi(X,Z) +
F(X,Z)\, dZ.\] Differentiating under the integral sign (which is
permitted by smoothness) gives
\[ \Psi_{,XX}=\int_0^Y -V_{,X}\Psi+V\Psi_{,X}+F_{,X}\, dZ.\]
This can be written as
\[ \Psi_{,XX}=\int_0^Y
[a(t)\Pi_{,t}+b(t)\Pi_{,x}+c(t)\Pi+d(t)e^{-3x}\mu_0(Z)]\exp(2(\frac{1-\nu}{1+\nu})x)\,dZ,\]
where in the integral it is understood that $t=t(X,Z)$ and
$x=x(X,Z)$. The term in the integrand in square brackets can be
bounded by an {\em a priori} term of the form in the statement of
the theorem; we use $M$ to represent such a term. Then
\[ |\Psi_{,XX}|\leq M\int_0^Y Z^{2(\frac{1-\nu}{1+\nu})}\,dZ,\]
wher we have used the definition $Y=te^x$ and absorbed a function of
type $a(t)$ into $M$. Evaluating the integral, we see that \[|
H^{-2}X^{\frac{\nu-1}{\nu}}e^{-x}\Psi_{,XX}|\leq a(t)M = M.\] In a
similar manner, we can show that
\[
|(1-\frac{\lam_1}{2}\alpha^\prime)H^{-2}X^{-\frac{1}{\nu}}e^{-x}\Psi_{,X}|\leq
M.\] Hence by (\ref{oddp0null2}), the theorem is proven for $i=0$.
The case $i=4$ is similar (but more straightforward).

For values of $\kp$ with $\kp\neq 0$, we simply point out that by
virtue of the self-similar and linear nature of the equation
(\ref{main}) and the linearity of the $\delta\Psi_i$ in $\Pi$ and
$\mu$, an identical argument to that above applies. \fin

We conclude by writing down the result describing bounds on the
perturbed Weyl scalars that ensues by considering data of the form
dealt with in Theorem 5. We omit the proof, noting that this
proceeds in exactly the way described in the summary proof of
Theorem 5: we apply Theorem 8 to a sequence of solutions generated
by data in $C^\infty_0$. Then the hypothesis that the limit of the
data exists and lies in (some) $H^{n,2}$ and the existence of the
{\em a priori} bounds in Theorem 8 ensures the existence of the
limits of those bounds.

\begin{theorem}\label{thm9}
Let $\kp\in[0,\kp_1)$ and let $\phi=e^{\kp x}\Pi$ and define
$\tmu(x)=e^{(\kp-3)x}\mu_0(x)=e^{(\kp-3)x}\mu(t_ie^x)$. Let $f\in
H^{3,2}(\mathbb{R})$, $g\in H^{2,2}(\mathbb{R})$, $\tmu\in
H^{2,2}(\mathbb{R})$. Then the perturbed Weyl scalars (\ref{oddp0})-
(\ref{oddp4}) calculated with respect to the unique solution
$\phi\in C([t_i,t_1],H^{1,2}(\mathbb{R}))$ of the initial value
problem consisting of (\ref{main}) with initial data
$\phi|_{\so}=f$, $\phi_{,t}|_\so=g$ satisfy the {\em a priori}
bounds (\ref{apbound}) of Theorem 8. In particular, the perturbed
Weyl scalars are finite at the Cauchy horizon $ch^+$. \fin
\end{theorem}

\section{The $l=1$ Perturbation.}
We return now to the $l=1$ perturbation. The treatment is
considerably more straightforward, but at the loss of full gauge
invariance of some of the results. The crucial difference for $l=1$
is that the metric perturbation quantity $k_A$ is no longer gauge
invariant. We find that under the infinitesmal transoformation
generated by $\bmt\xi=\xi S_\alpha dx^\alpha$, $k_A$ transforms as
\[ k_A\to k_A -r^2(r^{-2}\xi)_{,A}.\] Furthermore, the equation
(\ref{diveqn}) no longer holds. However the quantity $\Pi$ {\em is}
gauge invariant, and the equation (\ref{kvecredef}) holds. Since
$\ell=0$, this equation is readily solved once $L_A$ is determined.
This is done following the same procedure as for $l\geq 2$: we find
$L_A=\mu_0(v){\bar{\rho}}{\bar{k}}_A$. It is useful to take a
different approach (see the corresponding treatment of the
odd-parity $l=1$ perturbation in \cite{jm+carsten2}). The divergence
form of the conservation law (\ref{lincons}) indicates the existence
of a potential for $L_A$: we may write
$r^2L_A={\epsilon_A}^B\gamma_{,B}$. Comparison with the previous
version of $L_A$ shows that $\gamma=\gamma(v)$ with
$\gamma^\prime=-2\lam\mu_0$. The advantage of this is that we can
now write (\ref{kvecredef}) in the form
\[ {\epsilon_A}^B(16\pi\gamma-r^4\Pi)_{,B}=0,\]
yielding $r^4\Pi = 16\pi\gamma(v) + c,$ where $c$ is a constant of
integration. This form applies throughout the spacetime, including
the region $v<0$, where the background is flat. It is appropriate to
assume a vanishing matter perturbation ( $\gamma=0$) in this region,
and hence the appropriate boundary condition for $\Pi$ is to take
$c=0$. Thus the gauge invariant perturbation represented by $\Pi$ is
completely determined by the matter perturbation quantity $\gamma$
(or equivalently, $\mu_0$). Thus we have
\[ r^4\Pi=16\pi\gamma(v),\]
and there is no divergence at the Cauchy horizon (except possibly at
$r=0$, depending on the details of $\gamma$).

Consdering the perturbed Weyl scalars, we note that $\delta\Psi_0$
and $\delta\Psi_4$ vanish identically for $l=1$ (as expected: this
corresponds to the absence of dipole gravitational radiation).
$\delta\Psi_2$ is essentially $\Pi$, and so the comments above
regarding finiteness apply also to this Weyl scalar. From
(\ref{oddp1}) and (\ref{oddp2}), it is clear that $\delta\Psi_1$ and
$\delta\Psi_3$ are not gauge invariant for $l=1$. The effect of the
gauge transformation generated by $\bmt\xi$ is \by
\delta\Psi_1&\to&\delta\Psi_1
+4\frac{Q_1}{r}(r^{-2}\xi)_{,A}l^A,\\\delta\Psi_3&\to&\delta\Psi_3
+4\frac{{Q_1}^*}{r}(r^{-2}\xi)_{,A}n^A.\ey So while we cannot
ascribe any direct physical significance to these terms, it is also
true that any divergence of these quantities is gauge-dependent, and
can be removed by a gauge transformation. Thus the situation for
$l=1$ is the same as that for $l\geq 2$: an initially finite
perturbation remains finite at the Cauchy horizon.

\section{Conclusions}

In this paper, we have studied odd-parity perturbations of
self-similar Vaidya spacetime. More accurately, the study is of the
multipoles of the perturbation, i.e.\ the coefficients of the
(scalar, vector, tensor) spherical harmonics with respect to which
the perturbation quantities may be decomposed. The results are very
straightforward to state in rough terms: if the perturbation is
initially finite, then it remains finite as it impinges on the
Cauchy horizon. This statement of our results hides the details that
are represented by Theorems 1 - 9: the word `perturbation' refers to
the gauge invariant metric, matter and Weyl tensor quantities, and
`finite' refers both to integral energy measures and pointwise
values. Another detail is the meaning of the term `initial': we
slice the relevant region of spacetime using hypersurfaces generated
by the homothetic Killing vector field. These have the advantage of
being naturally aligned with the evolution of fields on the
spacetime in the sense that when we use this slicing and the
associated time coordinate $t$, the evolution equations are
independent of the space coordinate $x$. The disadvantage is that
these slices meet the (singular) scaling origin of the spacetime
rather than the regular axis. Consequently, one is driven to
consider data that vanish in a neighbourhood of this point. This is
undesirable, as this corresponds to data that are supported outside
the past light cone: this is clearly not the most general kind of
data one would like to consider. However, we can circumvent this
problem by studying a rescaled version of the fields, for example -
and most importantly - $\phi=r^\kp\Pi$. Then taking data with
$\phi\in H^{3,2}$ (and one derivative less for its time derivative,
along with an appropriate specification of the initial matter
perturbation), one obtains results for which the physical field
$\Pi$ {\em does not} necessarily vanish at the origin, and for which
all the finiteness results follow through.

In previous work, we considered the even parity perturbations of
self-similar Vaidya spacetime \cite{nolan-waters2}. Here, it was
necessary to take a Mellin transform of the (much more complicated)
system of perturbation equations. The results for the individual
modes of the perturbation were as for the perturbations studied
here: an  initially finite perturbation remains finite at the Cauchy
horizon. These results and those of the present paper provide
evidence for the stability of the naked singularity in self-similar
Vaidya spacetime. As noted in the introduction, this should not
however be considered a strong challenge to the cosmic censorship
hypothesis. Nonetheless, the results do indicate the propensity of
self-similar naked singularities to survive intact under linear
preturbations. Our hope is that the approach here can be applied to
cases of greater physical interest (especially those of perfect
fluid \cite{harada-maeda} and sigma model \cite{bizon-wasserman}
collapse) to yield insights into cosmic censorship in these cases.


\section*{References}

\begin{thebibliography}{odd-vaidya}
\bibitem{mtw} Misner, C.W., Thorne K.S. and Wheeler J.A. {\em
Gravitation} (W.H.\ Freeman and Co, New York, 1973).
\bibitem{wald} Wald, R.M. {\em General Relativity} (Univ.
Chicago Press, Chicago, 1984).
\bibitem{rendall} Rendall, A.D. {\em Class. Quantum Grav.} {\bf 9}, L99-L104 (1992).
\bibitem{christo} Christodoulou, D. {\em Ann. Math.} {\bf 149},
183(1999).
\bibitem{brady} Brady, P.R. {\em Prog. Theor. Phys. Suppl.}
{\bf 136}, 29 (1999).
\bibitem{carr-coley} Carr, B.J. and Coley A.A. {\em Gen.Rel.Grav.} {\bf 37} 2165
(2005).
\bibitem{nolan-waters1} Nolan, B.C. and Waters, T.J. {\em Phys.Rev.} {\bf
D66}, 104012 (2002).
\bibitem{nolan-scalar} Nolan, B.C. {\em Class. Quantum Grav.} {\bf 23}, 4523
(2006).
\bibitem{GS} Gerlach U.H. and Sengupta U.K. 1979 {\em
Phys. Rev.} {\bf D19} 2268
\bibitem{israel-poisson} Poisson, E. and Israel, W. {\em Phys. Rev.
Lett.} {\bf 63}, 1663 (1989).
\bibitem{jm+carsten2} Martin-Garcia J.M. and Gundlach C. 1999 {\em Phys.Rev.} {\bf D59} 064031
\bibitem{nolan-weyl} Nolan, B.C. {\em Phys. Rev.} {\bf D70}, 044004
(2004).
\bibitem{mcowen} McOwen, R.C. {\em Partial Differential
Equations: Methods and Applications} (Prentice Hall, New Jersey,
2003).
\bibitem{wald2} Wald, R.M. {\em J. Math. Phys.} {\bf 20}, 1056
(1979).
\bibitem{dafermos-mi} Dafermos, M. {\em Ann. Math.} {\bf 158}, 875
(2003).
\bibitem{stewart} Stewart, J.  {\em Advanced General Relativity}
(Cambridge University Press, Cambridge, 1991).
\bibitem{nolan-waters2} Nolan, B.C. and Waters, T.J. {\em Phys.Rev.} {\bf
D71}, 104030 (2005).
\bibitem{harada-maeda} Harada, T. and Maeda, H. {\em Phys. Rev.} {\bf
D63}, 084022 (2001).
\bibitem{bizon-wasserman} Bizon, P. and Wasserman, A. {\em Class. Quantum Grav.} {\bf 19}, 3309
(2002).
\end{thebibliography}
\end{document}